\documentclass[aps,prl,twocolumn,groupedaddress]{revtex4}
\pdfoutput=1
\usepackage{graphicx}
\usepackage{amsmath}
\usepackage{color}
\usepackage{natbib}
\usepackage{dcolumn}% Align table columns on decimal point
\usepackage{bm}
\long\def\symbolfootnote[#1]#2{\begingroup%
\def\thefootnote{\fnsymbol{footnote}}\footnote[#1]{#2}\endgroup} 

\begin{document}

\title{Breakdown of Long-Range Correlations in Heart Rate Fluctuations During Meditation}

\author{Nikitas Papasimakis\footnote[1]{Current Affiliation: Optoelectronics Research Centre, University of Southampton, Southampton SO17 1BJ, United Kingdom}\footnote[2]{Electronic Address: N.Papasimakis@orc.soton.ac.uk}}
\address{University of Athens, Faculty of Physics, Department of
Solid State Physics, Panepistimiopolis Zografou, 15784 Athens,
Greece}

\author{Fotini Pallikari}
\address{University of Athens, Faculty of Physics, Department of
Solid State Physics, Panepistimiopolis Zografou, 15784 Athens,
Greece}

\begin{abstract}The average wavelet coefficient method is applied to investigate the scaling features of heart rate
variability during meditation, a state of induced mental relaxation. While periodicity dominates the behavior
of the heart rate time series at short intervals, the meditation induced correlations in the signal become
significantly weaker at longer time scales. Further study of these correlations by means of an entropy analysis in the
natural time domain reveals that the induced mental relaxation introduces substantial loss of complexity at larger scales,
which indicates a change in the physiological mechanisms involved.
\end{abstract}

\maketitle

\section{INTRODUCTION}

In the past decade, the erratic behavior of heart interbeat
intervals in humans has attracted considerable attention and its
study has evolved to a virtually interdisciplinary topic,
encompassing diverse disciplines, such as cardiology, engineering
and more recently physics. It has been suggested that heart rate
variability (HRV) can provide non-invasive prognosis and diagnosis
tools for a number of pathological conditions \cite{peng93, peng97}.
Most importantly, it was demonstrated that heart rate variability exhibits various features that can be
used to distinguish between heart rate under healthy and
life-threatening conditions \cite{raab, ivanov99, kotani}.
Furthermore, it has been shown that the study of HRV can lead to
a better understanding of the dynamics of the underlying physiological
mechanisms \cite{struzik} in a variety of different conditions, such as apnea \cite{ivanov98}, sleep \cite{bunde} etc.
It is widely accepted that neuroautonomic control results in modulation of HRV \cite{parati} through the
antagonistic action of the two branches of the autonomic system, namely the sympathetic (SNS) and
parasympathetic (PNS). The first tends to increase heart rate and acts on long time scales (below 0.15 Hz,
while the latter is related to decreased heart rate and dominates frequencies above 0.15 Hz.  The effects of
meditation on heart rate were observed to excite prominent low frequency oscillations appearing to be driven by the
regularity of the imposed breathing pattern. This observation is rather unexpected considering that meditation
is thought of as a quiescent state of the brain. \cite{peng99}. Further studies suggested that different
meditative protocols induce different types of cardiovascular response that are not necessarily linked to
respiratory conditions (sinus arrhythmia) \cite{peng04}. In this work we investigate the behavior of heart rate
variability during
meditation at even longer time scales that range from roughly $15~sec$ to $20~min$ covering the low (LF) and very low (VL)
frequency band. By estimating the corresponding scaling exponent, we show
that meditation leads to the breakdown of long-range correlations
and loss of the 1/f character. Our findings extend previous studies \cite{fractal, sarkar} to indicate that complex
physiological changes take place during meditation.
\begin{figure}
\includegraphics[width=.45\textwidth]{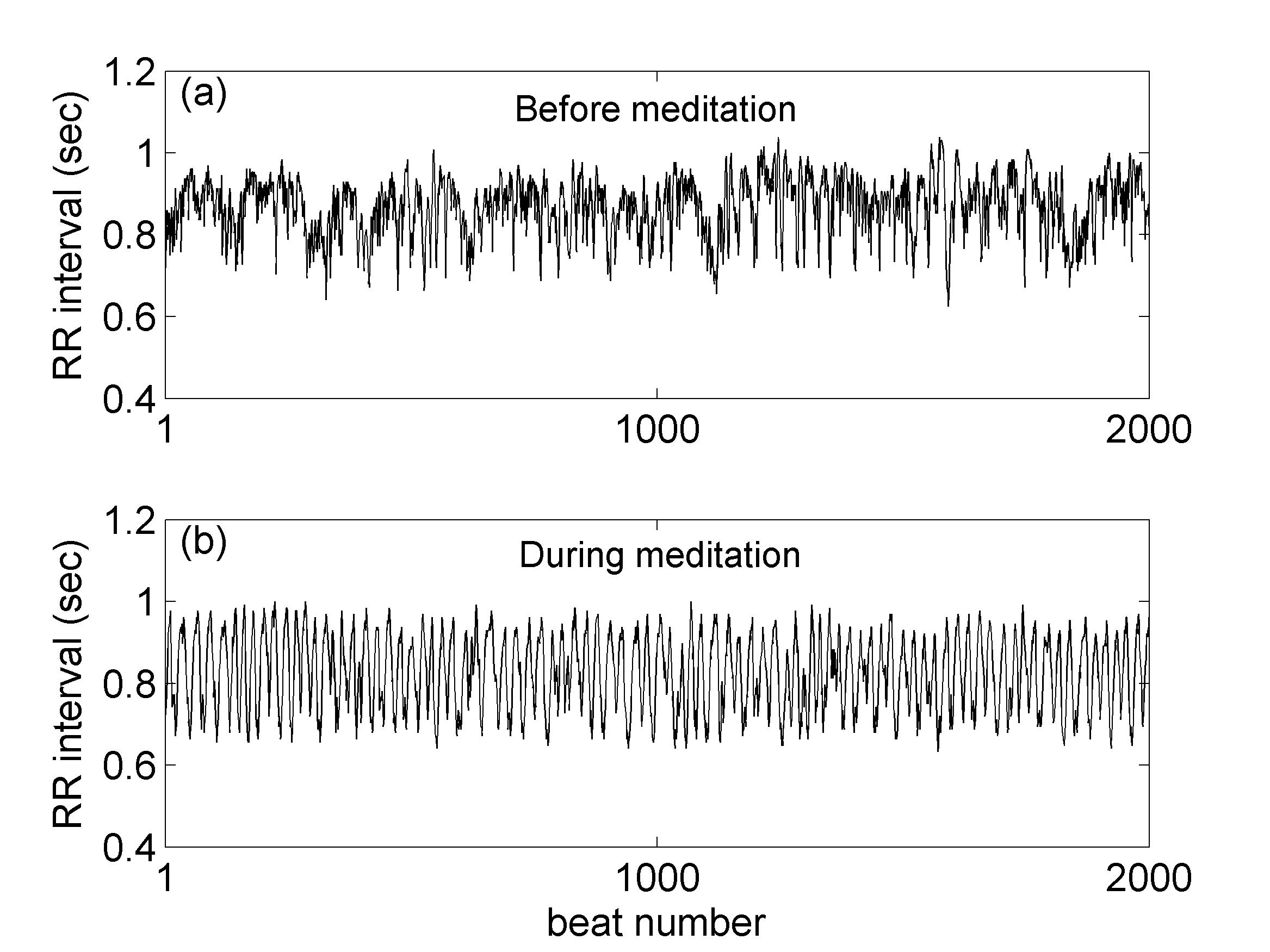}
\caption{\label{fig1} RR intervals before (a) and during (b)
meditation.}
\end{figure}

\section{METHODS}

We studied a publicly available (RR) interbeat interval database (www.physionet.org) consisting of data collected from
eight healthy subjects (aged 29-35) before and during
Qigong meditation (more information on the dataset and the meditation method can be found in
\cite{peng99}). The length of the time series varied between $50$ and $80$
minutes. One characteristic case of the RR interval time series, before and during meditation, is presented
in Figs.~1a \& 1b, respectively, where the evident irregular character of the time series before meditation
gives its place to smoother cyclic oscillations during meditation. The evolution of the time series was studied by the
application of two techniques, namely the average wavelet coefficient (AWC) method, which quantifies the
intensity of long-range correlations and the method that estimates the entropy in the natural time domain
which captures the complexity characteristics of the time series.

\subsection{Scaling exponent estimation - average wavelet coefficient method}

The complex features exhibited by HRV stem from the presence of the regulatory systems
operating at different time scales. Such systems that that generate fluctuations in the heart rate at all time
scales, can be characterized by means of a fractal analysis \cite{bunde}. A time series is termed fractal, when it possesses scale invariant characteristics at all scales. In practice
the fractal character is usually limited to a range of scales and is traditionally quantified by a scaling exponent $H$. When $0<H<1$, then the process possesses stationary increments and $H$ is usually called the Hurst exponent. In particular, a truly random process, $Y_H(t)$, with uncorrelated but stationary increments exhibits $H=0.5$.
For $H<0.5$ the process is mean-returning and is said to demonstrate
anti-persistent fractal behavior, while for $H>0.5$ the process is
mean-averting and persistent \cite{feder}. The graph of
$Y_{H}(t)$  is then a fractal, with fractal dimension $D=2-H$
\cite{feder,pallik}. The increments of $Y_H(t)$ also exhibit scale invariance with a scaling exponent equal to $H-1$, where $H$ is the scaling exponent of $Y_H(t)$. Therefore, it is through integration that one goes from a process of negative scaling exponents to one whose scaling exponents range in the interval (0,1). For differentiable processes
or processes with non-stationary increments $H>1$. Heart rate variability, on the other hand, has characteristics lying on the threshold of non-
stationarity for which $H\simeq1$, a behavior
typical for systems far from equilibrium. The power spectrum of such
a process scales with frequency $f$ as 1/f, known from this as 1/f
noise. In the present study, the analyzed time series exhibit close to zero or negative scaling exponents. In order to facilitate interpretation, they will always be considered to be the increments of the process of interest; hence all values of $H$ will refer to the latter and not to the original time series.

The estimation of the scaling exponent in the case of non-stationary
processes is a complex problem, especially when deterministic trends
are superimposed on the stochastic signal of interest. A range of
methods has been developed to overcome this
obstacle, each one being more appropriate for a specific kind of trends \cite{DFA,FDFA,simonsen}. A very
popular approach is the use of detrended fluctuation analysis (DFA). During DFA
the data are coarse-grained and a best-fit polynomial is
selected at each segment of the time series for every scale \cite{DFA}. While this
method is ideal for slowly varying polynomial trends, it fails when it comes to
cyclic components. Its shortcoming is illustrated in Fig.~2a, where DFA is applied on a random Gaussian
noise signal of standard deviation $\sigma=1$ with a superimposed sinusoidal trend of amplitude $1$ and
frequency $0.15~Hz$. DFA identifies correctly the global scaling behavior of the random signal yielding a
scaling exponent $H=0.51$ with standard error $0.03$. On the contrary, when the sinusoidal component is introduced, DFA
indicates
three different scaling regions each yielding a different scaling exponent and fails to isolate the behavior
of the stochastic component which might lead to spurious results \cite{struzik}. Measures to overcome such
difficulties have been addressed in the past \cite{Hu}.
\begin{figure}
\includegraphics[width=.45\textwidth]{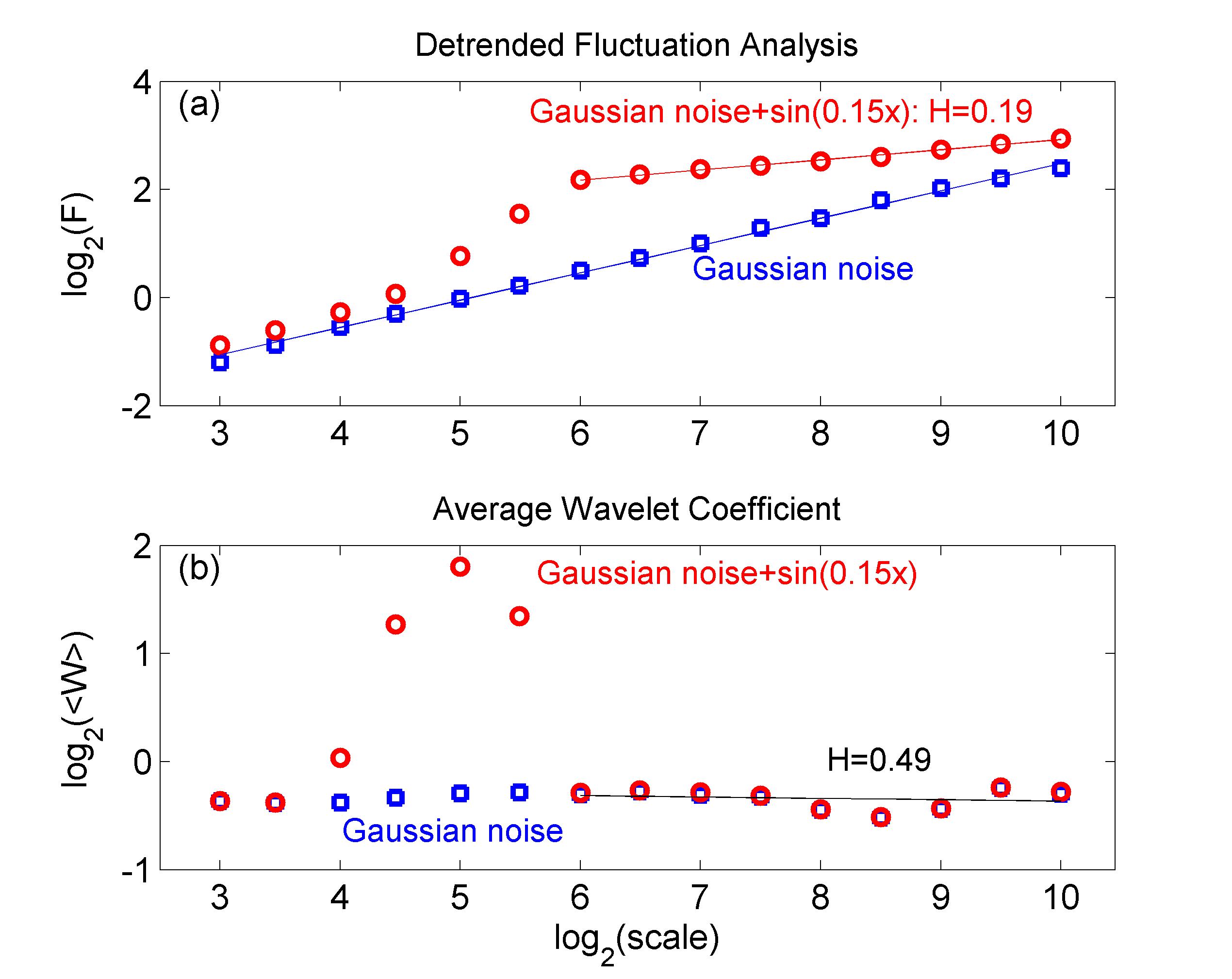}
\caption{\label{fig2} DFA (a) and AWC analysis (b) of Gaussian noise ($\sigma=1$) with (blue squares) and without (red circles) a superimposed sinusoidal trend ($sin(0.15x)$).}
\end{figure}

We have adopted here the wavelet-based method, since the wavelet transform
provides localization both in the time and frequency domain, while at the same time it can be made
orthogonal to polynomial trends. More precisely, the wavelet transform $W[f(t)]$ of a function $f(t)$ is
given by: $W_{a,b}[f(t)]=\int{f(t)\psi_{a,b}(t)^*dt}$, where $a$ and
$b$ are scale and location parameters, respectively, while the
"child" wavelet $\psi_{a,b}(t)$ is derived by appropriate scaling
of the so-called "mother" wavelet $\psi(t)$ through the transformation
$\psi_{a,b}(t)=\frac{1}{\sqrt{a}}\psi(\frac{t-b}{a})$
\cite{daubechies}. Figure 3 presents color maps of wavelet coefficients over the time-scale plane before and
during meditation. In both cases, the behavior of the interbeat intervals shows great complexity as
fluctuations of varying intensity occur at all scales. However, during meditation increased values of the wavelet coefficient
can be observed as a light colored band parallel to the horizontal (time) axis at scales in the range of $8-32$ beats. This is a signature of periodic components present at these
frequencies in agreement with previous studies \cite{peng97}. In our analysis we apply the average wavelet
coefficient method \cite{simonsen}. According to this approach, at
every scale $a$, the average value, $<W_a>$, of the wavelet
coefficients (over all values of the location parameter $b$) is
calculated. If the analyzed time series is self-similar with scaling exponent $H$, then its average wavelet coefficient will scale
as $a^{H+0.5}$ \cite{simonsen}. Therefore, one can estimate the scaling exponent of a
given time series by plotting the average wavelet coefficient versus
the scale parameter $a$ in a log-log plot (scalogram) and performing
a linear least-squares fit to the graph.
\begin{figure}
\includegraphics[width=.45\textwidth]{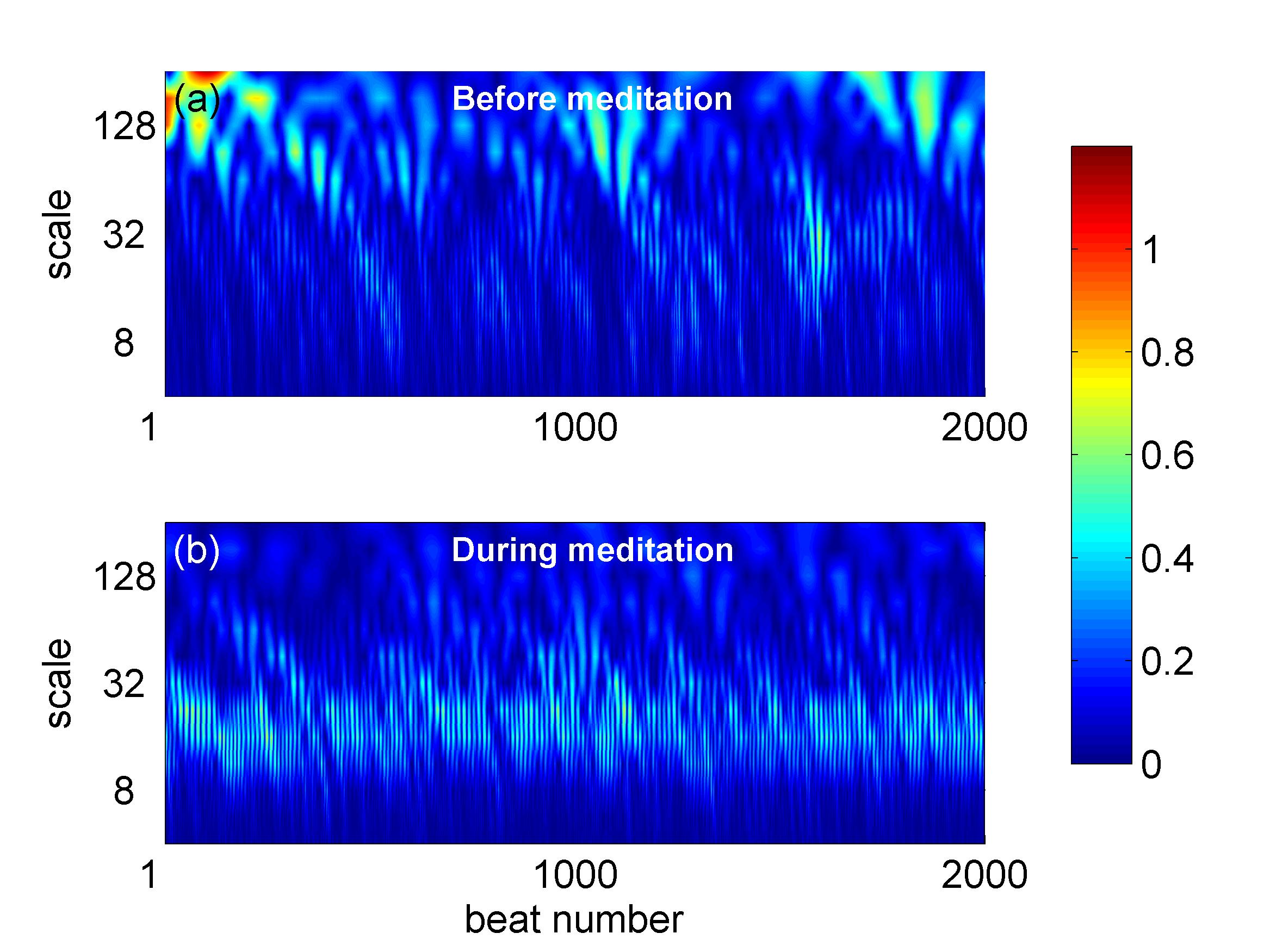}
\caption{\label{fig1} Absolute values of wavelet transform coefficients, $|W|$, of interbeat intervals for subject no. 1 of the studied dataset before (a) and during (b) meditation.}
\end{figure}

Figure 2b illustrates how the sinusoidal component can be isolated in the frequency (scale) domain by the
AWC method. Indeed, both the Gaussian noise and that with the superimposed sinusoidal trend are indistinguishable
and exhibit linear scaling apart from a small frequency region around the frequency of the sinusoidal
component,
where the scalogram shows an isolated peak indicative of the narrow frequency spectrum of the sinusoidal
component. These localization properties of the wavelet transform allow for a robust
estimation of the scaling exponent, even in the presence of strong cyclic components.

\subsection{Entropy in the natural time domain}

 The complexity characteristics of the heart rate time series are investigated by means of an entropy approach based on the concept of natural time $x
$, which has originally been introduced in \cite{varotsos02,varotsos03,varotsos05} to distinguish seismic electric signals \cite{varotsos82} from
artificial noise. This analysis was later extended \cite{varotsos04,varotsos05b} to electrocardiograms (ECG) \cite{varotsos03b}, since it is equally
applicable to deterministic as well as stochastic processes and physiological time series most likely contain components of both types (see \cite{
varotsos04, varotsos07} and references therein). The entropy approach in natural time works as follows.  In a signal composed of $N$ pulses, the
natural time $x$ is introduced by ascribing the value $x_m=m/N$ to the $m-th$ pulse, so that the analysis is made in terms of the couple $(x_m,Q_m)$,
where $Q_m$ denotes the duration of the $m-th$ pulse \cite{varotsos02}. The entropy $S$ is defined as \cite{varotsos03b} $S=<xlnx>-<x>ln<x>$, where $<
f(x)>=\sum_{m=1}^N\rho_m f(x_m)$ and $\rho_m=Q_m/\sum_{n=1}^N Q_n$. The entropy $S_-$ obtained after time reversal $T$ \cite{varotsos07b} so that $Tp_
m=p_{N-m+1}$, is different from the initial $S$ and their difference $\Delta S=S-S_-$ was found to be of major importance towards distinguishing
sudden cardiac death (SCD) individuals from healthy subjects (H) \cite{varotsos07}. In particular, the analysis considers a window of width $i$
sliding over the whole RR (beat-to-beat) interval time series, read in natural time, one pulse at a time. The entropies $S$, $S_-$ and their
difference $\Delta S_i$ are then estimated at every step. This difference provides a measure of the temporal asymmetry of the correlations in the
time series. More precisely, the physical meaning of  $\Delta S_i$ is derived in the following way. It has been shown \cite{varotsos06} (see also
Appendix $4$ of Ref. \cite{varotsos07}) that the estimated $\Delta S_i$   for the parametric family  $p(x,\epsilon)=1+\epsilon(x-1/2)$, $(|\epsilon|<1
)$ becomes negative for values of $\epsilon$  in the range $0<\epsilon<1$  (increasing trend). $\Delta S_i$  becomes positive for  $\epsilon<0$ (
decreasing trend). Applying this result in the case of ECGs indicates that the variation of the  $\Delta S_i$ with scale $i$ may be thought of as
capturing the net result of the competing mechanisms that either decrease or increase the heart rate (cf. the complex dynamics of heart rate is
attributed to the antagonistic activity of the parasympathetic and the sympathetic nervous systems decreasing and increasing heart rate,respectively
\cite{kotani}. Furthermore, the variability of $\Delta S_i$, quantified by its standard deviation $\sigma[\Delta S_i]$ with respect to time,
indicates whether one of the two mechanisms is dominant (large values $\sigma[\Delta S_i]$) or both are of  roughly the same intensity (small values $
\sigma[\Delta S_i]$). Such an analysis performed on $159$ ECG cases in natural time has shown \cite{varotsos07} that ventricular fibrillation starts
within three hours after $\Delta S_i$ (at the scale $i=13$ heartbeats) had become maximum for the $16$ out of $18$ SCD subjects. This time scale
corresponds to the so-called low frequency (LF) band found in heart rate (from $0.04$ to $0.15~Hz$), i.e. at around $0.1~Hz$, which is usually
attributed to "the process of slow regulation in blood pressure and heart rate" \cite{prokh}. Beyond the LF band, it was demonstrated that at scale i=
3 (HF band) a similar complexity measure based on natural time domain entropy allows for the distinction between SCD and in the healthy subjects,
since a smaller variability in entropy was observed in SCD than the healthy subjects, suggesting a breaking down of the high complexity \cite{
varotsos07}. In the present study, it is shown that the variance of the $\Delta S$ can be used to probe distinctive physiological changes occurring
at the VLF band during meditation, marking the loss of the 1/f character of heart rate at normal breathing conditions.

\section{RESULTS}

The AWC analysis was applied in the range of scales from $N_{min}=4$ up to $N_{max}=512$ conditioned by the prerequisite
that the
average  must be estimated over a sufficient number of uncorrelated wavelet coefficients even for $N_{max}$.
Color maps of the wavelet coefficients as a function of scale and time before and during meditation are
shown in Fig. 3 for one characteristic case. While before meditation, the wavelet transform exhibits complex
behavior over all scales, during meditation the wavelet presents strong maxima at a narrow scale region
around $N=15$ beats, where most of the signal power is concentrated. This variance of behavior is more clearly
seen in Fig. 4, where the corresponding scalograms are presented before (blue) and during
meditation (red) obtained by averaging the wavelet coefficients over the time axis. The time series before
meditation exhibits the typical 1/f behavior of HRV with linear scaling across all scales
and a scaling exponent of $1.1$ with standard error $0.1$. The picture changes dramatically during meditation. A strong
peak appears in
the scalogram at a frequency of about $0.22~Hz$, which marks the maximum of spectral power.  At longer time
scales, the graph returns to linear scaling, but with a much lower scaling exponent,
$H=0.43$ with standard error $0.1$, which indicates the weak anti-persistence of a mean-returning process.
\begin{figure}
\includegraphics[width=.45\textwidth]{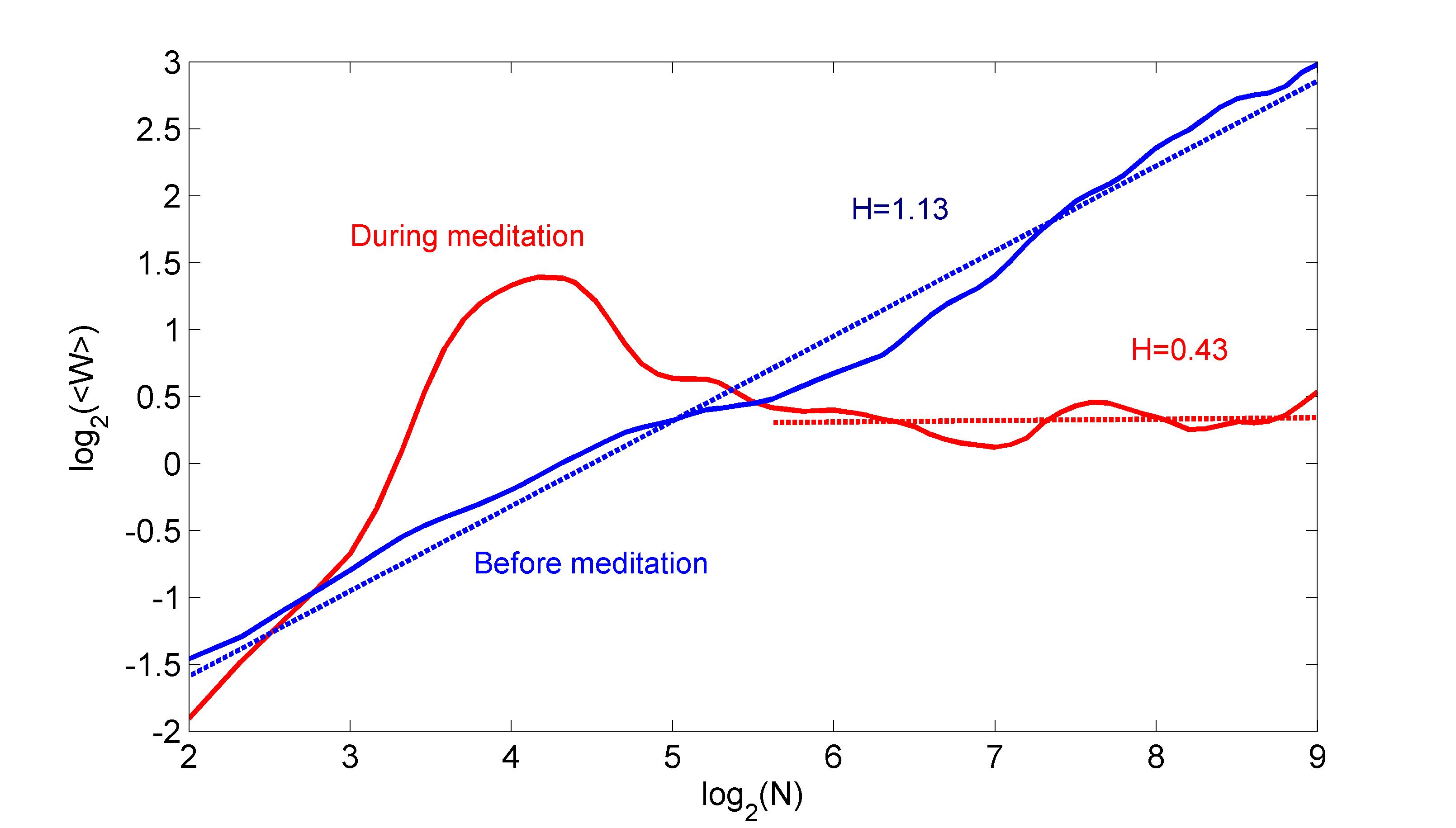}
\caption{\label{fig4} Characteristic scalogram before (blue) and during (red)
meditation for subject no. 1 of the studied dataset.}
\end{figure}

A summary of the results for all subjects is presented in Fig.~5. In all cases but two (6 and 7), the scaling
exponent during meditation drops significantly.
In the two other cases there is no statistically significant change in the scaling exponents
before and after meditation (at a $p=0.05$ level). The collective average scaling
exponent before meditation is $1.15$ with standard error $0.03$, while during meditation it
drops to $0.71$ with standard error $0.08$. This decrease indicates a change from the verge
of non-stationarity to a medium level of persistence. It implies that the overall long-range
correlations of the RR interbeat intervals at normal breathing have gotten significantly weaker
during meditation.  The $H$ values exhibit now a persistent fBm character \cite{feder,pallik} most likely induced by the regularity of the breathing
pattern.
\begin{figure}
\includegraphics[width=.45\textwidth]{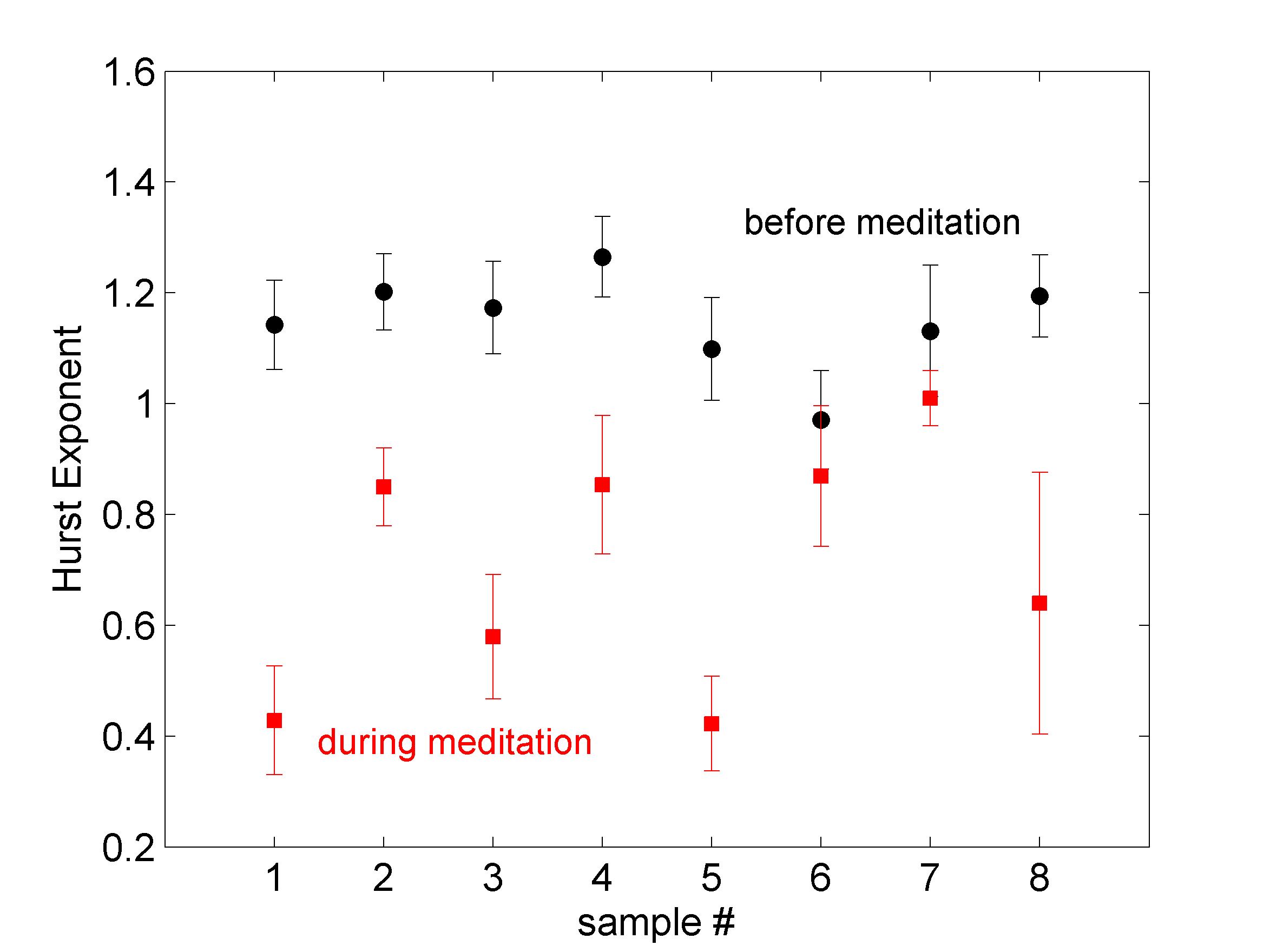}
\caption{\label{fig4} Scaling exponents before (blue) and during (red)
meditation.}
\end{figure}

The entropy $S$ in the natural time domain was calculated in a range of scales from $N_{min}=3$ to $N_{max}=60$ beats.
Then the time series were inverted with respect to time so that $S_-$ and consequently the difference $\Delta S$ was estimated. The standard deviation of $\Delta S$
as a function of scale (number of beats) is presented in Fig.~6 for a characteristic case (subject no. 1). In both conditions before
(blue) and during (red) meditation, $\sigma[\Delta S]$ increases rapidly at small scales up to about 10-15 beats.
At larger scales however, the two curves present distinctively different features. Before meditation, $\sigma[\Delta S]$
increases steadily with increasing scale but at a lower rate than before. On the other hand, during meditation, the situation
reverses and $\sigma[\Delta S]$ decreases with increasing scale. The decrease or increase of $\sigma[\Delta S]$ at longer time scales,
is quantified by fitting a linear trend and calculating the corresponding slope, $s$. Most other subjects' $\sigma[\Delta S]$ behaves similarly,
increasing before meditation and decreasing during meditation. An exception was seen for subject no. $7$,
where the two slopes are practically indistinguishable. Including this odd case, the average slope before meditation (over all subjects)
is found to be $14\cdot 10^{-6}$ with standard error $3\cdot 10^{-6}$, while during meditation its value is $-10\cdot 10^{-6}$ with standard error $4\cdot 10^{-6}$, implying that the two
values are statistically different.
\begin{figure}
\includegraphics[width=.45\textwidth]{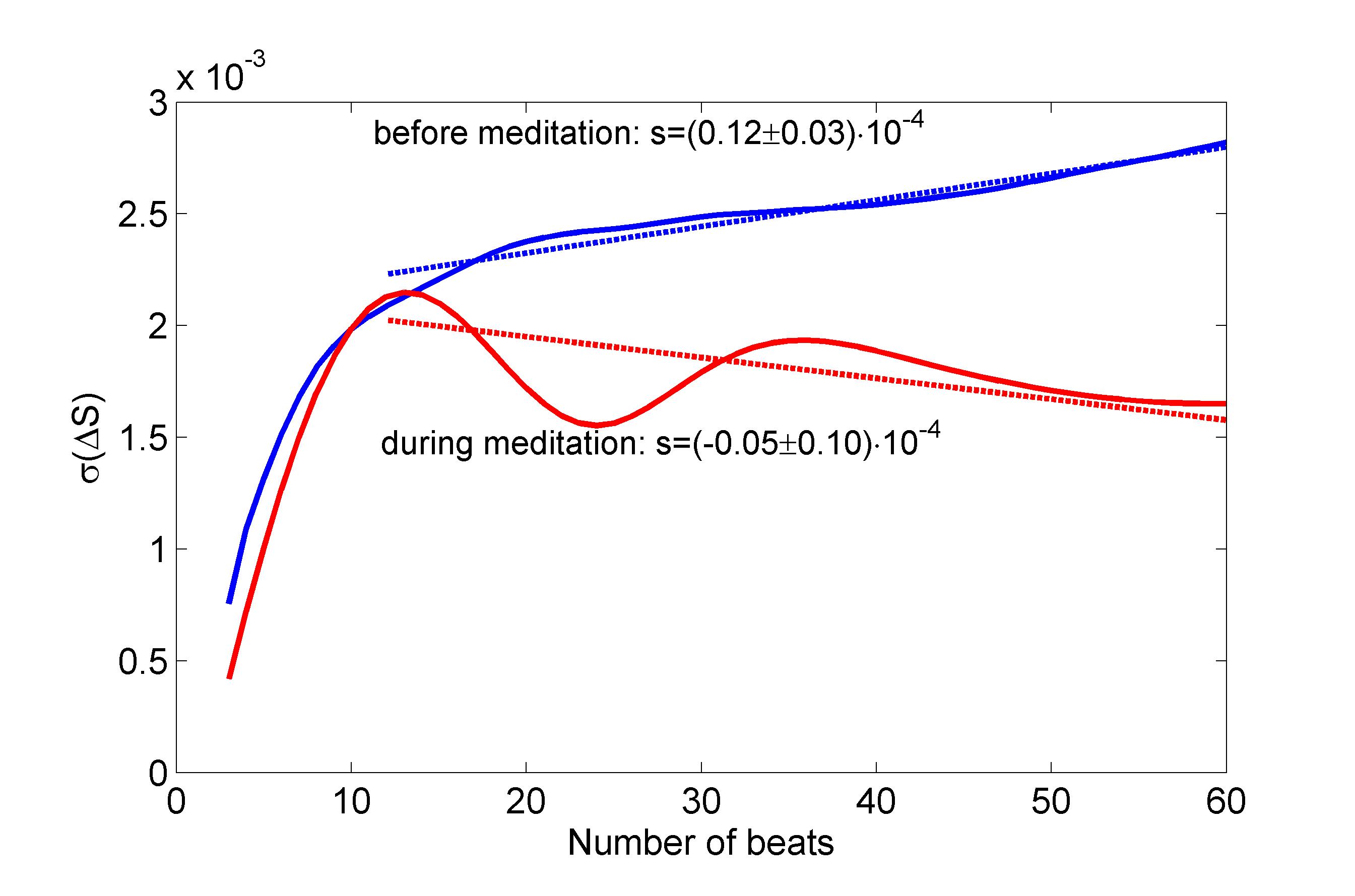}
\caption{\label{fig4} Standard deviation of $\Delta S$ before (blue) and (red) during meditation for subject no. 1 of the studied dataset. Dashed lines represent linear fits on the estimated values.}
\end{figure}

\section{DISCUSSION}

This study provides evidence that meditation, a state of induced deep mental relaxation, brings about changes in the cardiovascular system
 in addition to the previously observed enhancement of its low frequency components \cite{peng99}. Such changes occur at even lower frequencies
 and cannot be attributed to cardio-respiratory synchronization. In particular, the drop of the scaling exponent during meditation suggests that
 the $1/f$-noise character that was previously identified to portray infinite long-range correlations within the heart rate time series is now
 lost. That result can be related to the antagonistic control between the two branches of the autonomic nervous system which is responsible for
the HRV at the LF band \cite{struzik}. In particular, it appears that the heart rate interbeat
interval time series during meditation exhibits fBm character \cite{feder,pallik}, either appearing as weak persistence
or as anti-persistence. This diversity can be considered as evidence of an irregular competitive interplay
 between the sympathetic (SNS) and the parasympathetic nervous system (PNS). Whereas for $1/f$ noise large deviations from the average behavior
 of heart rate are common, such deviations become rare when persistence is observed ($1>H>0.5$). They get however quite sparse in the case of
 anti-persistence, i.e. when mean-returning traits are present in the heart rate time series.

 We argue here that during meditation the balance between SNS and PNS can be readily restored and consequently, small deviations of one of the two
 are quickly compensated by the other, avoiding thus the known large excursions from the typical heart rate behavior. Such a view can also
 explain the fact that the characteristic features are located even further at the VLF band. Indeed, although the physiological mechanisms responsible for HRV at this band
 have not been yet elucidated, it is believed that they include a strong contribution from the autonomic nervous system \cite{tforce}.

The results of the entropy analysis performed on the same data confirm the above picture. The standard deviation of $\Delta S$ during
meditation exhibits an overall decreasing trend at large scales, as opposed to its premeditation trend.  A diminishing trend in the
standard deviation of $\Delta S$ at larger scales, implies that the correlations in the heart rate do
not vary considerably during meditation. In fact, since $\Delta S$ quantifies the temporal asymmetry of these correlations, the observed
 behavior can be interpreted as follows: Under normal conditions, the asymmetry seen in the correlations of the heart
rate picture under
 time reversal, varies strongly and becomes more intense at larger scales \cite{varotsos08}. On the contrary, during meditation this variation
 is much weaker and it diminishes further with increasing scale. Hence, we can argue that during meditation the mechanisms responsible for the
 observed correlations (and their asymmetries) in heart rate variability do not vary considerably with respect to time.
Plausible causes for this change of behavior under meditation may involve both internal and external sources. One possible origin can be the
 calm, quiescent state adopted by those who meditate.  Another factor may be the shielding from external stimuli and the absence of physical
 movement, frequent in normal waken states that may no more pose great demands towards the balancing action between the two nervous systems.

It is inevitable to question the implications of the current results regarding the state of health of a subject who is meditating. There
is the assumption that the 1/f behavior is a sign of a healthy condition \cite{goldberger} and given that meditation seems to alter this
state of being, one may wonder if meditation is beneficial condition for the heart. This line of thinking, however, may lead to wrong assumptions,
 because an induced decrease in heart rate variability does not exclude its ability to behave in reverse fashion at normal conditions.
 In any case, further study comparing the heart rate variability of both meditating and non-meditating subjects under non-meditating conditions
 is needed in order to clarify whether the physiological effects of meditation are beneficial to the cardiovascular system.

\section{CONCLUSIONS}

In summary, we show that meditation induces distinct patterns in the heart interbeat intervals.  One is a periodic
feature in the LF band and another is the loss of the 1/f character in the VLF band.  Our findings are corroborated through the study of the entropy of the
time series  which reveals substantial loss of complexity in the LF and VLF bands. We argue that the observed behavior can be attributed,
at least partly, to changes in the balance between the sympathetic and parasympathetic branch of the autonomous nervous system.

\end{document}